\documentclass{elsarticle}

\usepackage{graphicx}
\usepackage{amssymb,latexsym,amsthm,amsmath} \usepackage{mathtools}
\usepackage{float}
\usepackage[draft]{changes}

\usepackage{hyperref}
\usepackage{color}
\usepackage{hyperref}
\journal{Journal of \LaTeX\ Templates}
\begin{document}

\begin{frontmatter}
\title{Statistical complexity and connectivity relationship in cultured neural networks}

\author[1,2]{A. Tlaie\corref{corrau}}
\cortext[corrau]{Corresponding author}
\ead{atboria@gmail.com }
\author[1,2]{L.M. Ballesteros-Esteban}
\author[1,2]{I.  Leyva}
\author[1,2]{I.  Sendi\~na-Nadal}
\address[1]{Complex Systems Group \& GISC, Universidad Rey Juan Carlos, 28933 M\'ostoles, Madrid, Spain}
\address[2]{Center for Biomedical Technology, Universidad Polit\'ecnica de Madrid, 28223 Pozuelo de Alarc\'on, Madrid, Spain}

\begin{abstract}
We explore the interplay between the topological relevance of a neuron and its dynamical traces in experimental cultured neuronal networks. We monitor the growth and development of these networks to characterise the evolution of their connectivity. Then, we explore the structure-dynamics relationship by simulating a biophysically plausible dynamical model on top of each networks' nodes. In the weakly coupling regime, the statistical complexity of each single node dynamics is found to be anti-correlated with their degree centrality, with nodes of higher degree displaying lower complexity levels. Our results imply that it is possible to infer the degree distribution of the network connectivity only from individual dynamical measurements. 
\end{abstract}

\begin{keyword}
\texttt{Complex Networks, Neuron Models, Network Inference, Cultured Networks}
\end{keyword}

\end{frontmatter}

\section{Introduction}

One of the main research lines in the study of the dynamics of complex networks has been the deep relationship between the connectivity and the dynamics of the nodes, and how this interaction shapes the emergence of a collective state such as synchronisation \cite{Pecora1998, Barahona2002, Boccaletti2006, Arenas2008}. An enormous effort has been devoted to the understanding of this phenomenon, and the knowledge gathered so far has driven the advance in crucial applications in brain dynamics \cite{Bullmore2009}, power grids \cite{Rohden2012}, and in many other fields where synchronisation is {essential \cite{Pluchino2005,Fujiwara2011}} for the system's proper functioning. 

Commonly, studies have focused in states of full synchronisation \cite{Arenas2008}. Nevertheless, there are very relevant cases in which only a partial or weak synchronisation level is achieved \cite{Rodriguez1999,Lachaux1999,Pecora2014}, and often this state becomes optimal to balance functional integration and segregation in the system \cite{Tononi1994,Rad2012,Sporns2013} while a complete coordination is evidencing the existence of a   pathological condition. 

Several investigations \cite{Gomez-Gardenes2007,Li2008,Navas2015} have shown that nodes play different roles in the ensemble dynamics depending on their topological position and intrinsic dynamics \cite{Skardal2014}. One of the most explored situation is that of the hubs acting as coordinators of the dynamics of the whole system \cite{Heuvel2013,Papo2014, Zamora-Lopez2016, Deco2017}, being the first nodes to synchronise among them  \cite{Pereira2010} and to the mean field \cite{Zhou2006}, while the rest of the nodes progressively locks the hubs dynamics. 

 This effect of the topology on the dynamics in the weakly synchronised regime  opens the question of whether it is possible to infer the network architecture from statistical correlations among the coupled units  \cite{Gomez-Gardenes2007,Arenas2006}.  Currently, a great amount of the research is being conducted in this sense. In particular, the computational neuroscience field roots in the hypothesis that dynamical correlations (which can be recorded in non-invasive ways) are greatly constrained and induced by the anatomical structure of the brain \cite{Bullmore2009}. From  these site-to-site correlation maps, the {\it functional brain networks}, it is often possible to obtain information about the underlying topological networks \cite{Honey2009}.

However, it has been less explored the fact that this structural-dynamical interaction also plays in the other way around: just as the dynamics of each node influences the ensemble, the ensemble imprints its structural marks into the dynamics of each individual node \cite{Pereira2010,Zhou2006}. We make the assumption that, long before the coupling strength is high enough to induce synchronisation, the dynamical changes at the node level are encoding the imprint of its structural role. This relevant feature could be used to extract information about the network without making any reference to pairwise correlations, particularly in those cases where the structure is unknown or unreliable, as we showed in a previous work \cite{Tlaie2018}. 

Here, we extend our study of the influence that the ensemble has over the node dynamics to an experimental case. We culture networks of neurons coming from \textit{Schistocerca gregaria} and study the potential relationship between a simulated dynamical model (the Morris-Lecar neuron) and the anatomical network structure of a neuronal culture. The main motivation for this study is that in cultured neuronal networks the simultaneous obtention of structural and dynamical information is not possible, either because one recording technique influences the other measurement or mainly because the culture is not able to survive to both measurements. 

\section{Experimental setup: culturing the network}
\label{sec:experiment}

To analyse the spatial structure of a real network, we focus on the study of cultured neuronal networks (CNNs), considered as a  simplified version of a more complex network of the central nervous system \cite{Eckman2012, Fuchs2007,Vanpelt2005}. For this purpose , we analyse the network structure in this CNN model by means of optical microscopy techniques, extracting the detailed connectivity and its statistical topological properties.

Our CNNs were obtained from \textit{Schistocerca gregaria} specimens, also known as desert locusts. As they share basic neuronal features with vertebrates, this invertebrate model has been recently used in neuroscience as an easier approach for the understanding of more  complex neural systems \cite{Burrows1996}. The large size of its neurons makes it ideal for observing the structure of the network, as an alternative to  the mammals. 

In our experiments we follow the protocol described in \cite{Anava2013}. Each locust is dissected to extract its frontal ganglion,   formed by approximately 100 neurons \cite{Ayali2002}. To obtain an intermediate neuron density that allows us to study a complex network morphology we extracted $12$ ganglia per culture. After the dissection, the frontal ganglia endure a chemical and mechanical procedure to remove all the connections and dissociate the neurons. The neuronal somata are cultured in a Petri dish, in an enriched environment to allow the neurites to regrowth and form a new connectivity network. The cultures are monitored \textit{in vitro} from day 0 (DIV0, DIV=days in vitro) to day 14 (DIV14). The data used in this work correspond to $6$ cultures grown in the same conditions. We inspected the morphological features of the cultured networks using a phase contrast inverted microscope (Eclipse Ti-S, Nikon) with a $10$x air objective (Achromat, ADL, NA $0.25$) and an automated motorised $XYZ$ stage controller. High resolution images were obtained in a daily basis.

In order to analyse the spatial network, we need to extract the corresponding mathematical graph. To do so, we process the culture images by means of an image segmentation algorithm \cite{Santos-Sierra2014,Santos-Sierra2015}. In Fig. \ref{fig:Segmentation_Full} we portray the whole process, starting from a typical microscope image of the culture (Fig.~\ref{fig:Segmentation_Full}(a) shows just a small area) and ending up with the output of the segmentation detecting neurons (and aggregates of neurons) and the neuronal processess connecting them Fig.~\ref{fig:Segmentation_Full}(a). The algorithm is summarised in the following steps:

\begin{figure}[t]
  \centering
  \includegraphics[width=1.\textwidth]{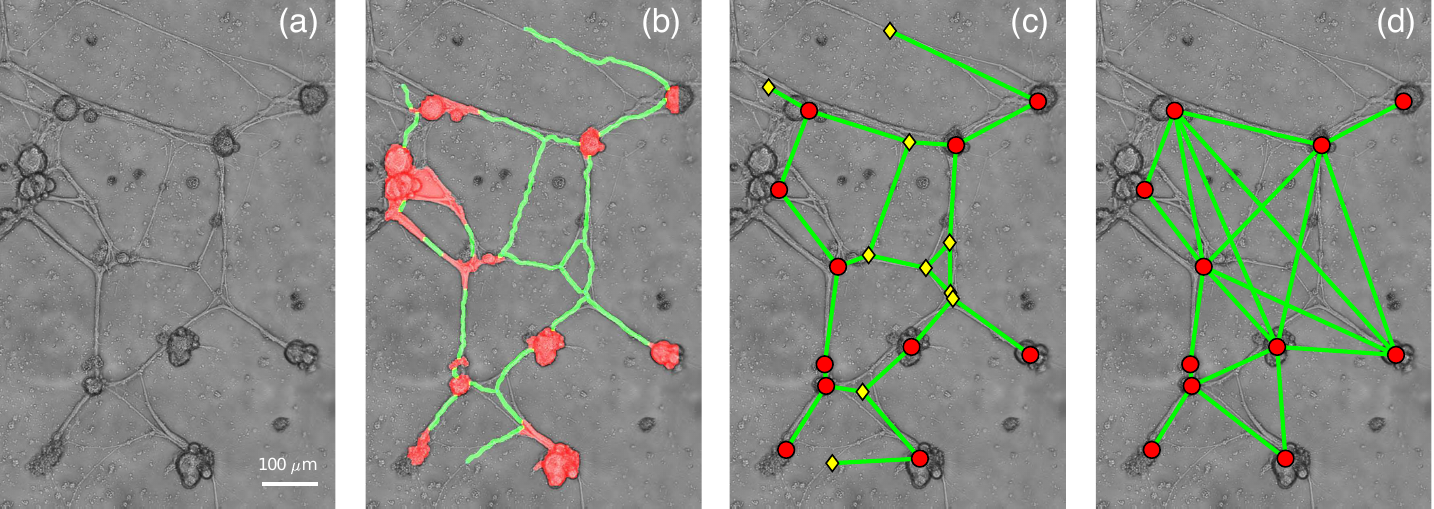}
    \caption{Image segmentation processing steps and extraction of the network graph. (a) Red layer of a RGB cut of a culture $6$ DIV old. (b) Output of the segmentation algorithm of the region of interest. Single neurons and aggregates of neurons are highlighted in red while the  neurites are marked in green. (c) Mapping of the segmentation objects in (b) into a full graph where circles (neurons and neuronal clusters) and diamonds (branching and end points of neurites, with no neurons) are the nodes, these being connected with green lines, representing the neurites.  (d) Graph representing the projection of the full graph into the cluster graph  with only neuronal clusters (red) and neurites (green), where the links represent the existence of a path between cluster nodes through junction nodes in the full graph in (c). }\label{fig:Segmentation_Full}
\end{figure}
\begin{enumerate}

\item The red layer of a RGB  high-resolution image of the recorded cultured network  is processed (Fig.\ref{fig:Segmentation_Full}(a)).

\item The image is segmented and thresholded to separate background from foreground areas. Then neurons and aggregates of neurons (red areas in Fig.~\ref{fig:Segmentation_Full}(b)) and neurites  (green paths in Fig.~\ref{fig:Segmentation_Full}(b))  are identified separately .

\item  Both neurons and neurites are connected and coded in the adjacency matrix where single and clustered neurons are the nodes and neurites are the links between them. Branching and end points of the neurites  are also registered as junction nodes in the graph, even when there is no neuron on them. This provides a complete version of the graph that we call the {\it full graph} (Fig. \ref{fig:Segmentation_Full}(c)) with two types of nodes, those corresponding to neurons and the ones denoting a branching point in the neuronal process path connecting two neurons.

\item  The previous data is used to build a reduced version of the graph, where only neurons (or neuronal clusters) are the nodes, and the links represent the existence of a path between neuronal clusters, eventually through branching points. In this graph, junction  nodes have been removed and we observe a more direct path between neuronal clusters, obtaining a simple version of the matrix, {\it the cluster graph} (Fig.
\ref{fig:Segmentation_Full}(d)).
\end{enumerate}

\begin{figure}
  \centering
  \includegraphics[width=1.\columnwidth]{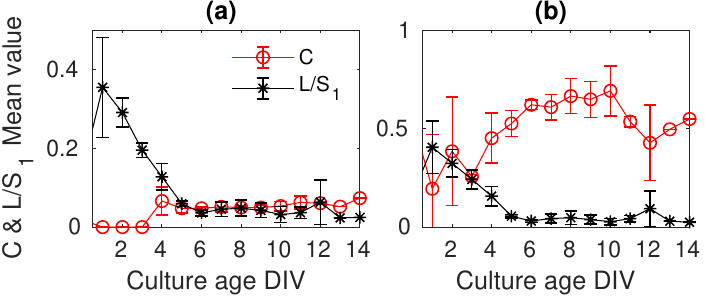}
    \caption{Results of the cultured network analysis. Mean value of the clustering coefficient $C$ and mean path length $L$ normalised  by the size of  the largest connected component $S_1$ (a) in a full graph, where clusters and junctions are the nodes, and (b) in a cluster graph, where the only nodes are clusters of neurons. Each point is the average over 6 experiments.}\label{fig:Culture}
\end{figure}

We analyse the morphological and topological properties of the cultured network using both full and cluster graphs, where the links are unweighted. With the purpose of characterising the segregation and integration of the cultured neuronal network along  the experiment, in Fig. \ref{fig:Culture} we measure the longitudinal progression of the averaged clustering coefficient ($C$) and shortest path length $L$, normalised by the size of the largest connected component $S_1$ \cite{Boccaletti2006}, resulting $L/S_1$. The relationship between $C$ and $L$ is often used as an indicator of the balance between the local and long-distance connectivity in the network. These aforementioned parameters were measured in both the full graph (with neuronal clusters and junction nodes) and in the cluster graph (with only neuronal clusters nodes).

In the full graph (Fig. \ref{fig:Culture} (a)), the normalised shortest path $L /S_1$ shows a high mean value at the early days of the culturing, where the connectivity is still not fully developed. Between DIV3 and DIV6 there is a significant decrease and showed no significant change thereafter, meaning a high integration degree in the mature culture network. The clustering coefficient was characterised by a very low mean value, showing a slight increase between DIV3 and DIV6, when the network development occurs. The low mean values of $C$ are due to the fact that both neurons and branching points are considered nodes meaning that the probability of forming triangles is reduced (see Fig.~\ref{fig:Segmentation_Full}(b)). 

In the case of the cluster graphs (Fig. \ref{fig:Culture} (d)) we observe a similar trend in $L /S_1$ (see Fig.~\ref{fig:Culture}(b)) as the one described for the full graph. On the contrary, $C$ exhibits higher mean values, with a more acute increase between DIV 3 and DIV 6, that coincides with the most intense developmental phase. As neuronal aggregates are the only nodes in this cluster graph and junctions are not represented as nodes, the mean values of $C$ are more accurate with the connected structure. After that point, in the mature neuronal network these two parameters keep constant values \cite{Santos-Sierra2014,Santos-Sierra2015}. 

The analysis of these parameters in both types of graphs concludes with the emergence of a mature cultured neuronal network from an initial random stage. This evolved structure is characterised by  high clustering coefficients and low mean path values, indicating the presence of  a mature network with  high segregation (favored by high clustering values) and integration (facilitated by the existence of small shortest paths) levels. These are the characteristics of a small world structure, where the high tendency to form clusters of nodes in highly interconnected subgroups and short distance between them contribute to an optimal functionality in the network \cite{Santos-Sierra2014}.

\begin{figure}
  \centering
  \includegraphics[width=0.5\textwidth]{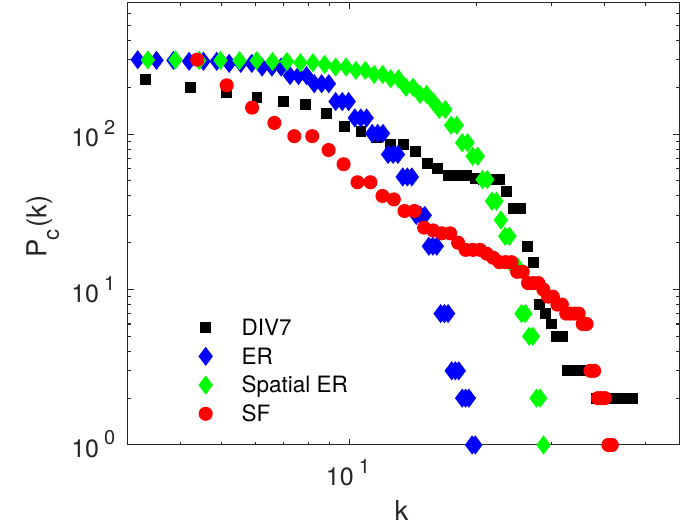}
    \caption{Comparative plot of the cumulative degree distribution $P_c(k)$ of an experimental clustered neuronal network at DIV7 (black squares) and equivalent synthetic networks: random Erd\"os-Renyi (ER, blue diamonds), scale-free obtained with the Barab\'asi-Albert algorithm (SF, red dots) and a spatial network where distance-dependence linkage pattern (spatial-ER green diamonds). We see that the cultured network's distribution shows a similar behavior, for low degrees, as the SF one, and there is a region in which it is similar to the spatial network (namely, the decay for large values of $k$ is almost identical).}\label{fig:Deg_dist}
\end{figure}

We also analized the degree distribution $P(k)$ of these networks, being $k$ the number of links that each node has. In Fig.~\ref{fig:Deg_dist} we plot an example of the cumulative degree distribution $P_c(k)$ of an in-vitro clustered network at DIV7 (black squares), compared to equivalent simulated networks (same number of nodes and links) obtained from usual generative models: random Erd\"os-Renyi (ER, blue diamonds), scale-free obtained by Barabasi-Albert algorithm (SF, red dots) and a spatial network with a distance-dependence linkage pattern (spatial-ER green diamonds). As described in\cite{Santos-Sierra2014}, the cultured networks belong to the single-scale type as they show a well defined $k$. The study of $P_{cum}(k)$ reveals a fast decay with a large number of nodes with similar number of connections, and a few ones with a different and large node degrees. As it can be seen, the experimental connectivity largely differs from the pure random ER, showing instead shared features between SF and spatial networks. 

\section{Dynamical model}

Once we have extracted the connectivity of real neuronal cultures , we can provide a dynamical behavior to their nodes, in order to enable the exploration of the potential interplay between structure and dynamics.  For this study we implemented the bio-inspired Morris-Lecar (ML) model \cite{morris1981,Tlaie2018}, whose equations describing the membrane potential behavior for each unit read \cite{Sancristobal2013, Navas2015}:  

\begin{align}\label{ML}
C \dot{V_i} &=  -{}\overbrace{g_{\rm X} M_{\infty}( V_i-V_{\rm X} )}^{\text{Ionic channels}} + q\xi_i + \underbrace{\frac{\sigma}{K} \sum_{j} a_{ij} \overbrace{e^{-2(t-t_j)} (V_0-V_i)}^{\text{Synaptic function}}}_{I_i} + I^{ext}_i ,  \\\nonumber
\dot{W_i} &= \phi \,\tau_W  ( W_{\infty}-W_i ) 
\end{align}

\noindent where $V_i$ and $W_i$ are, respectively, the membrane potential and the fraction of open $\rm K^+$ channels of the $i$th neuron and $M_{\infty}, W_{\infty}$, and $\tau_W$ are hyperbolic functions dependent on $V_i$ and $\phi$ is a reference frequency. The parameters $g_{\rm X}$ and $V_{\rm X}$ account for the electric conductance and equilibrium potentials of the $\rm X=\{K,Ca,\text{leaky}\}$ channels. The external current $I_i^{ext}=50.0$ mA is the same for all the neurons and is chosen such that neurons are sub-threshold to neuronal firing which is induced by the white Gaussian noise $q\xi_i$ of zero mean and intensity $q$. The coupling of the neuron $i$th with the neuron ensemble is described by the injected synaptic current $I_i$, given by the superposition of all the post-synaptic potentials emitted by the neighbours of node $i$ in the past, being $t_j$ the time of the last spike of node $j$, and the corresponding element of the adyacency matrix  is $a_{ij}=1$ if there is a link between nodes $i,j$ and $a_{ij}=0$ otherwise. The synaptic conductance $\sigma$, normalised by the largest node degree present in the network $K$, plays the role of coupling intensity.  

Additionally, the channel voltage-dependent saturation values are given by the following functions:    
\begin{eqnarray}\nonumber
M_{\infty}(V_i) &=& \frac{1}{2}\left[1+\tanh\left(\frac{V_i-V_1}{V_2}\right)\right], \\
W_{\infty}(V_i) &=& \frac{1}{2}\left[1+\tanh\left(\frac{V_i-V_3}{V_4}\right)\right], \\\nonumber
\tau_W(V_i) &=& \cosh \left( \frac{V_i-V_3}{2V_4} \right).
\end{eqnarray}

We chose the parameters such that the neurons in the simulations corresponded to type II class excitability for the neuron dynamics, which means that a discontinuous transition is found in the dependence of the spiking frequency on the external current.  The values for all the parameters can be found in Refs.~\cite{Tlaie2018,Leyva2011}.

We have to remark, however, that this model was originally conceived for a single neuron and in this work we are dealing with aggregates of $ ~20 $ of them as our individual nodes. Although this quantity is not enough for employing a \textit{neural mass model}, it should be noted that, strictly, it is not a single neuron either.

\section{Statistical characterisation of the dynamics}

With the purpose of providing a solid description of the system, we use two different quantities: a global and a local one. The global measure is the \textit{synchronisation level} of the network, i.e. how similar are the dynamical outputs of our units, while the local one is an individual measure of the \textit{complexity} of a node's time series.

\subsection{Synchronization measure}

In order to quantify the level of synchronization we estimate how many neurons fire within the same time window. The total simulation time $T$ is divided in $n=1,\dots,N_b$ bins of a convenient size $\tau$, such that $T=N_b\tau$, and the binary quantity $B_i(n)$ is defined such that $B_i(n)=1$ if the $i$th neuron spiked within $n$th interval and $0$ otherwise. The synchronisation between the spiking sequences of neurons $i$ and $j$ is therefore characterised with the pairwise correlation matrix  $s_{ij}\in [0,1]$
\begin{equation}
s_{ij}=\frac{\sum_{n=1}^{N_b}B_i(n)B_j(n)}{\sum_{n=1}^{N_b}B_i(n)\sum_{n=1}^{N_b}B_j(n)},
\end{equation}
where the term in the denominator is a normalisation factor and $s_{ij}=1$ means full coincidence between the two spiking series. The ensemble average of $s_{ij}$, $S=\langle s_{ij}\rangle =\frac{2}{N(N-1)}\sum_{i,j=1, i\neq j}^N s_{ij}$ is a measure of  the global synchronisation in the network.  In Fig.~\ref{fig:ResultadosCultivos}(a) we plot an example of the averaged value of $S$ as a function of $\sigma$ for an experimental CNN clustered network with $N=246$ nodes, obtained at DIV7 as explained in Sec. \ref{sec:experiment}.  A transition from an asynchronous to an almost synchronous firing is observed as the synaptic conductance $\sigma$ is increased, which confirms that the structure is suitable for inducing synchronisation in the system. 

\subsection{Statistical complexity}

Once we know the effect that the relationship between network topology and dynamics has on the global state, we explore the effect that the presence of the ensemble has on the single node dynamics by measuring the {\it statistical complexity} of the single nodes along the synchronisation process.   As the typical neuronal dynamics exhibited by Eq.~(\ref{ML}) consists of a sequence of $L$ spikes whose amplitude variability is negligible,  we focused on the complexity $C_i$ of the sequence of inter-spike intervals $(t_l-t_{l-1})$ (ISI) of each neuron. 

\begin{figure}
  \includegraphics[width=\textwidth]{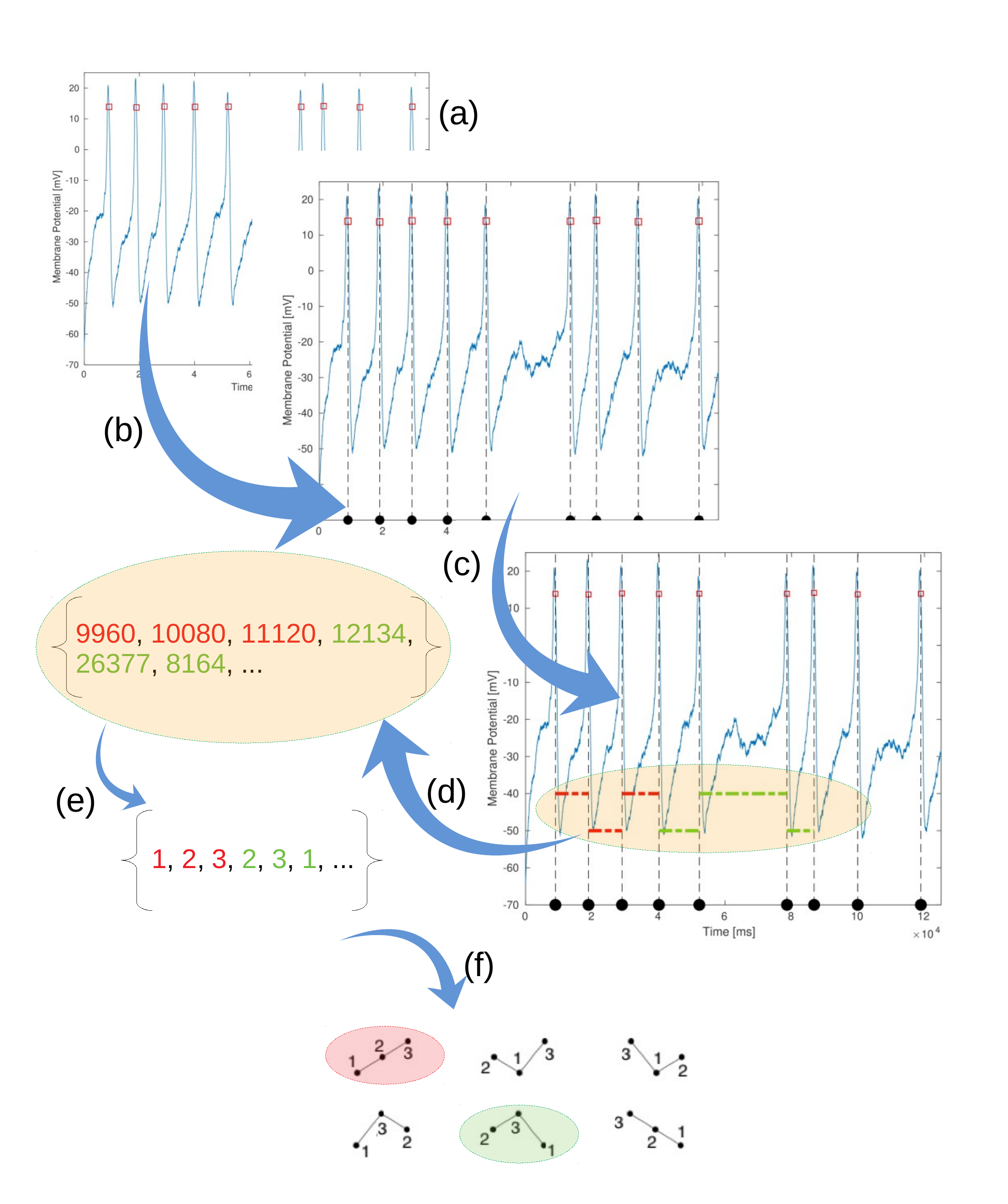}
  \caption{Illustration of the Ordinal Patterns formalism. (a) Detection of  the maxima of the simulated (Morris-Lecar neuron) signal. (b) Extraction of the timestamps at which these maxima have been attained. Consecutive timestamps are subtracted (c) and their differences (ISIs) are stored in an array (d). The time series of ISIs is divided in sequences of length $D$ ($D = 3$ in this example), and consecutive points in each sequence are compared and classified based on their relative values   (e). Each sequence of length $D$  is given a natural number and the probability of the realization of each possible $D$-symbols sequence is computed (f).  The probabilities are then used to construct the complexity measure.}\label{fig:OP_Process}
\end{figure}

The ordinal patterns formalism  \cite{Bandt2002} associates a symbolic sequence to a series \cite{Letellier2008}, transforming the actual values of the data series into a set of natural numbers. To do that, the  ISI series of each neuron is divided in sequences of length $D$. In each sequence, the data values are ordered in terms of their relative   magnitudes \cite{Rad2012}, which provides the corresponding symbolic sequence. The information content of these sequences is then evaluated as a function of the complexity measure. The complete process is illustrated in Fig. \ref{fig:OP_Process}.

This is a broad-field, well-established and known method, statistically reliable and robust to noise, extremely fast in computation and with a clear definition and interpretation in physical terms. It is derived from two also well-established measures (divergence and entropy), also easily interpretable when analysing non linear dynamical systems. In addition, it only requires soft criteria, namely that the time series must be weakly-stationary, i.e., for $k\le D$, the probability for $\rm{ISI}_t<\rm{ISI}_{t+k}$ should not depend on time \cite{Bandt2002}, and that $M>>D!$ (where $M$ is the number of points of the entire time series of ISIs), which are easily checkable. We proceed in the following way, as shown in Fig.~\ref{fig:OP_Process}:

\begin{itemize}

\item From each single node time series in the simulated (Morris-Lecar neuron) signal, we detect the spikes (a) and extract the duration between two consecutive spikes  (b). 

\item We compute series of  the inter-spike time intervals  ISI (c) and save them in an array (d), which will be our object of study. 

\item The  ISI series is divided in sequences of length $D$ ($D = 3$ in this illustration, (e)). We compare consecutive points in each sequence and associate a natural number to each of them (f), ranking them based on their relative size. 

\item We count how many times a certain symbolic  sequence (or {\it pattern}) $\pi$ of length $D$ appears ($N_\pi$).

\item Then, we  define a probability of occurrence for each pattern: $P_\pi = \frac{N_\pi}{N_T}$, where $N_T$ is the total number of sequences of length $D$ in which  the time series is divided, i.e. $N_T = (L-1)/D$, being $L$ the total number of spikes.

\item We construct a \textit{ probability distribution}, which we call $P$ from now on, from all possible symbolic sequences of length D with probability $P_\pi$.
\end{itemize}

Once the probability distribution $P$ is obtained, the statistical complexity is defined. It is a measure that should be minimal both for pure noise and absolute regularity, and provide a bounded value for other regimes. Being this so, we need to characterise the disorder and a correcting term (i.e., a way of comparing known probability distributions with the actual one). The statistical complexity ($C$),  as defined in Ref. \cite{Martin2003}, is the product of the Permutation Entropy ($H$) and the Disequilibrium ($Q$).

To define the permutation entropy $H$, the first step is the evaluation of the Shannon entropy, that gives an idea of the \textit{predictability} of the series:
\begin{equation}
S[P] = - \sum_{j=1}^{D!} p_j \cdot \log(p_j)
\end{equation}
The permutation entropy corresponds to the normalisation of $S$ with respect to the entropy of the uniform probability distribution, $S_{max}$:
\begin{align}
&H = \frac{S}{S_{max}}, \quad S_{max} = S[P_e], \\
& P_e \equiv \lbrace p_i=1/D! \rbrace_{i=1,...,D!}  \implies 0 \leq H \leq 1 \nonumber
\end{align}

Regarding the disequilibrium $Q$, it is a way of measuring the distance of the actual probability distribution $P$ with the equilibrium probability distribution $P_e$. This notion of {\it distance} can be acquired by several means; in this work, we adopt the statistical distance given by the Kullback-Leibler \cite{Kullback1951} relative entropy ($K$):
\begin{align}\label{eq:KL} \nonumber
K[P|P_e] &= - \sum_{j=1}^{D!} p_j \cdot \log(p_e) + \sum_{j=1}^{D!} p_j \cdot \log(p_j) = \\
&= S[P|P_e] - S[P] 
\end{align}

\noindent where $S[P|P_e]$ is the Shannon cross entropy. If we now symmetrise Eq.~(\ref{eq:KL}), we get the Jensen-Shannon divergence ($J$):

\begin{equation}\label{eq:J}
J[P|P_e] = (K[P|P_e]+K[P_e|P])/2 \underbracket{\rightarrow}_{(*)} J[P|P_e] = S[(P+P_e)/2] - S[P]/2 - S[P_e]/2
\end{equation}

\noindent where $(*)$ is simply the rewritten version in terms of $S$. Finally, we can write the disequilibrium $Q$ as the normalised version of $J$ as:
\begin{equation}
Q = Q_0 J[P|P_e]
\end{equation}
with $Q_0 = {\frac{N+1}{N}\log(N+1)-2\log(2N)+\log(N)}^{-1}$, implying again $0 \leq Q \leq 1$. We then just have to multiply $H$ and $Q$ to obtain the \textit{Complexity measure}:

\begin{equation}
C = H \cdot Q
\end{equation}

\section{Results}

\begin{figure}
  \centering
  \includegraphics[width=0.75\textwidth]{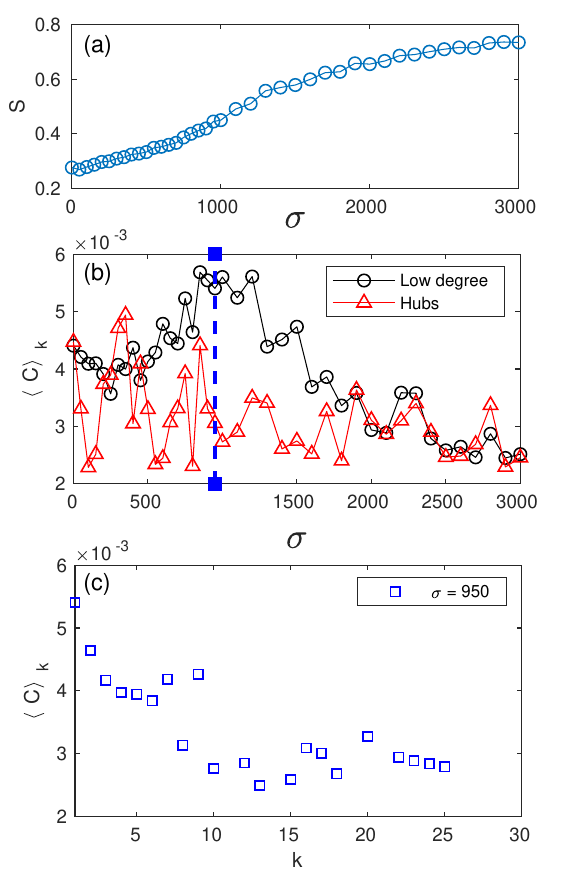}
    \caption{Dependence of the statistical complexity at the node level and its topological role in an experimental neuronal network 7 days in vitro old with $N=246$ Morris-Lecar neurons. (a) Synchronisation curve for the explored values of the synaptic conductance. (b) Complexity values $\langle C \rangle_k$ vs. $d$ for low ($k=3$) and high ($k=20$) degree node values. (b) $\langle C \rangle_k$  vs. $k$ for the conductance value ($\sigma = 950$) marked in (b) with a vertical dashed line. Each point is the average of $6$ network realisations.}
\label{fig:ResultadosCultivos}
\end{figure}

We summarise our results for the statistical complexity in a cultured neuronal network in Fig.~\ref{fig:ResultadosCultivos}. As commented above, in  panel (a) we show  as a reference the  synchronisation level vs the synaptic conductance $\sigma$ for the dynamics simulated on top of a DIV7 experimental network. In panel Fig.~\ref{fig:ResultadosCultivos}(b) we plot the value of $\langle C\rangle_k$ as a function of the conductance $\sigma$ for two nodes with  high ($k=30$) and poor ($k=3$) connectivity, being $\langle C\rangle_k =\sum_{[i|k_i=k]} C_i/N_k$, with $N_k$ the number of nodes with degree $k$.  The results evidences that, on that same route to synchronisation, there exists differences between how hubs and peripheral nodes behave due to the presence of the ensemble, even when the global synchronisation level is still very low.  

The main detail that catches our attention is that the peripheral nodes show a greater complexity than the hubs ($\sigma = 950$). To further explore this finding, in the third panel we depict the statistical complexity vs the degree, for this value of $\sigma$. We can extract an interesting result here: for $\sigma = 950$, \textit{there exists an anti-correlation between $\langle C\rangle_k$ and $k$.}

This anti-correlation observed in cultured neuronal cultures is not as evident as the one reported in Ref.~\cite{Tlaie2018} in SF networks, but taking into account that the structure, which has grown in a limited spatial domain, does not belong to the class of power-law networks (as it was already discussed) and that the model (Morris-Lecar) was not originally designed for this kind of neuronal aggregates, one can conclude that the anti-correlation between the statistical complexity $C$ and the degree $k$ is a quite robust feature. 

\section{Conclusions}

In Ref.~\cite{Tlaie2018} we investigated  the relationship between the statistical complexity and topology
 in  synthetically generated networks. Here, we focused on the study of real-world topologies, as the ones exhibited by self-organised neuronal cultures. The longitudinal study of the morphology of these networks shows an evolution in the topology from isolated neurons to a percolated heterogeneous topology with small-work properties. 

 In order to study the structure-dynamics interaction in these networks, we simulated a dynamical model (the Morris-Lecar neuron) on top of  experimental neuronal networks at the mature developmental stage.  We evidenced that, in the weakly coupled regime, it is possible to anti-correlate the individual node statistical complexity of the series of the neuronal inter-spike intervals with the degree of a node. Therefore, it would be possible to infer the degree distribution of the network from node dynamical measurements, which confirms the result obtained for synthetic networks \cite{Tlaie2018}. This approach based on the computation of complexity values retrieved from single node dynamics, provides a different perspective than the usual methods of network inference, since it does not imply node-to-node calculations. Additionally, our method does not impose the need of measuring the dynamics of every node: it can be an incomplete measure, an still it will provide their relative roles. We hope this approach will be useful in applications where the knowledge of the degree distribution, instead of the detailed connectome, provides a sufficient insight over an unknown topology and about the functioning of the underlying system.

\section{Acknowledgements}

Financial support  from the Ministerio de Econom\'ia y Competitividad of Spain under project FIS2017-84151-P and from the Group of Research Excelence URJC-Banco de Santander is acknowledged. A.T. and L.B.-E.  acknowledge support from the European Youth Employment Initiative.

\bibliography{references}

\begin{thebibliography}{10}
\expandafter\ifx\csname url\endcsname\relax
  \def\url#1{\texttt{#1}}\fi
\expandafter\ifx\csname urlprefix\endcsname\relax\def\urlprefix{URL }\fi
\expandafter\ifx\csname href\endcsname\relax
  \def\href#1#2{#2} \def\path#1{#1}\fi

\bibitem{Pecora1998}
L.~M. Pecora, {Synchronization conditions and desynchronizing patterns in
  coupled limit-cycle and chaotic systems}, Phys. Rev. E 58~(1) (1998)
  347--360.

\bibitem{Barahona2002}
M.~Barahona, L.~M. Pecora, Synchronization in small-world systems, Phys. Rev.
  Lett. 89 (2002) 054101.

\bibitem{Boccaletti2006}
S.~Boccaletti, V.~Latora, Y.~Moreno, M.~Chavez, D.-U. Hwang, {Complex networks:
  Structure and dynamics}, Phys. Rep. 424~(4–5) (2006) 175--308.

\bibitem{Arenas2008}
A.~Arenas, A.~D\'iaz-Guilera, J.~Kurths, Y.~Moreno, C.~Zhou, Synchronization in
  complex networks, Phys. Rep. 469~(3) (2008) 93 -- 153.

\bibitem{Bullmore2009}
E.~Bullmore, O.~Sporns, {Complex brain networks: graph theoretical analysis of
  structural and functional systems}, Nat. Rev. Neurosci. 10~(3) (2009)
  186--198.

\bibitem{Rohden2012}
M.~Rohden, A.~Sorge, M.~Timme, D.~Witthaut, Self-organized synchronization in
  decentralized power grids, Phys. Rev. Lett. 109 (2012) 064101.

\bibitem{Pluchino2005}
A.~Pluchino, V.~Latora, A.~Rapisarda, Changing opinions in a changing world: a
  new perspective in sociophysics, Int. J. Mod. Phys. C 16~(04) (2005)
  515--531.

\bibitem{Fujiwara2011}
N.~Fujiwara, J.~Kurths, A.~D\'{\i}az-Guilera, Synchronization in networks of
  mobile oscillators, Phys. Rev. E 83 (2011) 025101.

\bibitem{Rodriguez1999}
E.~Rodriguez, N.~George, J.~P. Lachaux, J.~Martinerie, B.~Renault, F.~J.
  Varela, Perception's shadow: Long-distance synchronization of human brain
  activity, Nature 397~(6718) (1999) 430--433.

\bibitem{Lachaux1999}
L.~Jean-Philippe, R.~Eugenio, M.~Jacques, V.~F. J., Measuring phase synchrony
  in brain signals, Human Brain Mapping 8~(4) (1999) 194--208.

\bibitem{Pecora2014}
L.~M. Pecora, F.~Sorrentino, A.~M. Hagerstrom, T.~E. Murphy, R.~Roy, Cluster
  synchronization and isolated desynchronization in complex networks with
  symmetries, Nature Communications 5 (2014) 4079.

\bibitem{Tononi1994}
G.~Tononi, O.~Sporns, G.~M. Edelman, {A measure for brain complexity: relating
  functional segregation and integration in the nervous system.}, Proc. Natl.
  Acad. Sci. 91~(11) (1994) 5033--5037.
\newblock \href {https://doi.org/10.1073/pnas.91.11.5033}
  {\path{doi:10.1073/pnas.91.11.5033}}.

\bibitem{Rad2012}
A.~A. Rad, I.~Sendi\~na Nadal, D.~Papo, M.~Zanin, J.~M. Buld\'u, F.~del Pozo,
  S.~Boccaletti, Topological measure locating the effective crossover between
  segregation and integration in a modular network, Phys. Rev. Lett. 108 (2012)
  228701.

\bibitem{Sporns2013}
O.~Sporns, {Network attributes for segregation and integration in the human
  brain}, Curr. Opinion Neurobiol. 23 (2013) 162--171.
\newblock \href {https://doi.org/10.1016/j.conb.2012.11.015}
  {\path{doi:10.1016/j.conb.2012.11.015}}.

\bibitem{Gomez-Gardenes2007}
J.~G{\'{o}}mez-Garde{\~{n}}es, Y.~Moreno, A.~Arenas, {Paths to synchronization
  on complex networks}, Phys. Rev. Lett. 98~(3) (2007) 034101.
\newblock \href {https://doi.org/10.1103/PhysRevLett.98.034101}
  {\path{doi:10.1103/PhysRevLett.98.034101}}.

\bibitem{Li2008}
D.~Li, I.~Leyva, J.~A. Almendral, I.~Sendi\~na Nadal, J.~M. Buld\'u, S.~Havlin,
  S.~Boccaletti, Synchronization interfaces and overlapping communities in
  complex networks, Phys. Rev. Lett. 101 (2008) 168701.

\bibitem{Navas2015}
A.~Navas, J.~A. Villacorta-Atienza, I.~Leyva, J.~A. Almendral, I.~Sendi\~na
  Nadal, S.~Boccaletti, Effective centrality and explosive synchronization in
  complex networks, Phys. Rev. E 92 (2015) 062820.

\bibitem{Skardal2014}
P.~S. Skardal, D.~Taylor, J.~Sun, Optimal synchronization of complex networks,
  Phys. Rev. Lett. 113~(14) (2014) 144101.

\bibitem{Heuvel2013}
M.~P. {Van Den Heuvel}, O.~Sporns, {Network hubs in the human brain}, Trends
  Cogn. Sci. 17 (2013) 683--696.

\bibitem{Papo2014}
D.~Papo, M.~Zanin, J.~A. Pineda-Pardo, S.~Boccaletti, J.~M. Buldu, J.~M.
  Buld{\'{u}}, {Functional brain networks: great expectations, hard times and
  the big leap forward}, Phil. Trans. R. Soc. B 369~(1653) (2014)
  20130525--20130525.

\bibitem{Zamora-Lopez2016}
G.~Zamora-L{\'{o}}pez, Y.~Chen, G.~Deco, M.~L. Kringelbach, C.~Zhou,
  {Functional complexity emerging from anatomical constraints in the brain: the
  significance of network modularity and rich-clubs}, Sci. Rep. 6~(1) (2016)
  38424.

\bibitem{Deco2017}
G.~Deco, T.~J. {Van Hartevelt}, H.~M. Fernandes, A.~Stevner, M.~L. Kringelbach,
  {The most relevant human brain regions for functional connectivity: Evidence
  for a dynamical workspace of binding nodes from whole-brain computational
  modelling}, NeuroImage 146 (2017) 197--210.

\bibitem{Pereira2010}
T.~Pereira, Hub synchronization in scale-free networks, Phys. Rev. E 82 (2010)
  036201.

\bibitem{Zhou2006}
C.~Zhou, J.~Kurths, {Hierarchical synchronization in complex networks with
  heterogeneous degrees}, Chaos 16~(1) (2006) 015104.

\bibitem{Arenas2006}
A.~Arenas, A.~D\'{\i}az-Guilera, C.~J. P\'erez-Vicente, Synchronization reveals
  topological scales in complex networks, Phys. Rev. Lett. 96 (2006) 114102.

\bibitem{Honey2009}
C.~J. Honey, O.~Sporns, L.~Cammoun, X.~Gigandet, J.~P. Thiran, R.~Meuli,
  P.~Hagmann, Predicting human resting-state functional connectivity from
  structural connectivity, Proceedings of the National Academy of Sciences
  106~(6) (2009) 2035--2040.

\bibitem{Tlaie2018}
A.~Tlaie, I.~Leyva, R.~Sevilla-Escoboza, V.~Vera-Avila, I.~Sendina-Nadal,
  Dynamical complexity as a proxy for the network degree distribution, arXiv
  preprint arXiv:1807.09629 (2018).

\bibitem{Eckman2012}
J.-P. Eckmann, L.~Feinerman, O.and~Gruendlinger, E.~Moses, J.~Soriano,
  T.~Tlusty, The physics of living neural networks, Physics Reports 449 (2007)
  54--76.

\bibitem{Fuchs2007}
E.~Fuchs, A.~Ayali, A.~Robinson, E.~Hulata, E.~Ben-Jacob, Coemergence of
  regularity and complexity during neural network development, Dev. Neurobiol.
  67 (2007) 1802--1814.

\bibitem{Vanpelt2005}
J.~van Pelt, I.~Vadja, P.~S. Wolters, M.~Corner, G.~J.~A. Ramakers, Dynamics
  and plasticity in developing neuronal networks in vitro, Prog. Brain Res. 147
  (2005) 173--188.

\bibitem{Burrows1996}
M.~Burrows, The Neurobiology of an Insect Brain, Oxford University Press, 1996.

\bibitem{Anava2013}
S.~Anava, I.~Saad, A.~Ayali, The role of gap junction proteins in the
  development of neural network functional topology, Molecular Insect Biology
  22 (2013) 457--472.

\bibitem{Ayali2002}
A.~Ayali, Y.~Zilberstein, C.~N., The role of gap junction proteins in the
  development of neural network functional topology, J. Exp. Biol. 205 (2002)
  2825--2832.

\bibitem{Santos-Sierra2014}
D.~de~Santos-Sierra, I.~Sendi\~na Nadal, I.~Leyva, J.~Almendral, S.~Anava,
  A.~Ayali, D.~Papo, S.~Boccaletti, Emergence of small-world anatomical
  networks in self-organizing clustered neuronal cultures, PLoS ONE 9 (2014)
  E85828.

\bibitem{Santos-Sierra2015}
D.~de~Santos-Sierra, I.~Sendi\~na Nadal, I.~Leyva, J.~Almendral, S.~Ayali,
  A.Anava, C.~S\'anchez-\'Avila, S.~Boccaletti, Graph-based unsupervised
  segmentation algorithm for cultured neuronal networks' structure
  characterization and modeling, Citometry A 87 (2015) 513--523.

\bibitem{morris1981}
C.~Morris, H.~Lecar, Voltage oscillations in the barnacle giant muscle fiber,
  Biophys. J. 35~(1) (1981) 193--213.

\bibitem{Sancristobal2013}
B.~Sancrist{\'{o}}bal, R.~Vicente, J.~M. Sancho, J.~Garc\'ia-Ojalvo, {Emergent
  bimodal firing patterns implement different encoding strategies during
  gamma-band oscillations.}, Front. Comput. Neurosci. 7 (2013) 18.

\bibitem{Leyva2011}
I.~Leyva, A.~Navas, I.~Sendi{\~n}a-Nadal, J.~M. Buldu, J.~A. Almendral,
  S.~Boccaletti, Synchronization waves in geometric networks, Physical Review E
  84~(6) (2011) 065101.

\bibitem{Bandt2002}
C.~Bandt, B.~Pompe, Permutation entropy: A natural complexity measure for time
  series, Phys. Rev. Lett. 88 (2002) 174102.

\bibitem{Letellier2008}
C.~Letellier, Symbolic sequence analysis using approximated partition, Chaos,
  Solitons \& Fractals 36~(1) (2008) 32 -- 41.
\newblock \href {https://doi.org/https://doi.org/10.1016/j.chaos.2006.06.025}
  {\path{doi:https://doi.org/10.1016/j.chaos.2006.06.025}}.

\bibitem{Martin2003}
M.~Martin, A.~Plastino, O.~Rosso, Statistical complexity and disequilibrium,
  Physics Letters A 311~(2-3) (2003) 126--132.

\bibitem{Kullback1951}
S.~Kullback, R.~A. Leibler, On information and sufficiency, The annals of
  mathematical statistics 22~(1) (1951) 79--86.

\end{thebibliography}

\end{document}